\documentclass[prl,aps,twocolumn,showpacs,floatfix]{revtex4}
\usepackage{graphicx}
\usepackage{bm}
\usepackage{amsmath}
\usepackage{amssymb}
\usepackage{amsfonts}
\usepackage{dcolumn}%
\usepackage{bm}
\newcommand{\be}{\begin{equation}}
\newcommand{\ee}{\end{equation}}
\newcommand{\bea}{\begin{eqnarray}}
\newcommand{\eea}{\end{eqnarray}}

\let\oldphi\phi
\let\phi\varphi
\let\varphi\oldphi

\begin{document}
 \title{Evidence of Raleigh-Hertz surface waves and shear stiffness anomaly in granular media}
\author{L. Bonneau}
\author{B. Andreotti}
\author{E. Cl\'ement}
\affiliation{PMMH, ESPCI, CNRS (UMR 7636) and Univ. Paris 6 \& Paris 7,\\ 10 rue Vauquelin, 75005 Paris France.}
 \date{\today}

\begin{abstract}
Due to the non-linearity of Hertzian contacts, the speed of sound in granular matter increases with pressure. Under gravity, the non-linear elastic description predicts that acoustic propagation is only possible through surface modes, called  Rayleigh-Hertz modes and guided by the index gradient. Here we directly evidence these modes in a controlled laboratory experiment and use them to probe the elastic properties of a granular packing under vanishing confining pressure. The shape and the dispersion relation of both transverse and sagittal modes are compared to the prediction of non-linear elasticity that includes finite size effects. This allows to test the existence of a shear stiffness anomaly close to the jamming transition.
\end{abstract}

\maketitle

Many disordered condensed systems --~mostly multi-stable systems such as regular fluids trapped in a glassy phase~\cite{TWLB02}~-- can loose their shear rigidity, characteristic of ordinary solids, when they are on the verge to yield. For static, non-cohesive granular assemblies, the external pressure $p$ is the only source of confinement that may jam the packing in the rigid phase. Thus, at zero thermal agitation, a solid/fluid transition would be reached under a vanishing confining pressure $p$ i.e. when the packing becomes a marginal solid with just the minimal amount of contacts $\mathcal{Z}$ per grain suited to sustain a large scale elastic network, i.e. at isostaticity ($\mathcal{Z}=\mathcal{Z}_{\rm iso}$). In the simple case of spherical, friction-less, disordered granular packing~\cite{OHSLN03}, it has been shown theoretically that the local linear response becomes non-affine near jamming, due to the overwhelming appearance of weak floppy modes of divergent spatial extension~\cite{WSNW05}. These soft modes make the shear to bulk modulus ratio vanish at the transition. Can  this anomalous linear response be evidenced experimentally? Is this jamming scenario robust enough to describe real granular matter, including frictional contacts~\cite{SHESS07, MRJWM08}, gravity loading or non-spherical particles, close to jamming? Does it provide a generic picture for the glassy transition of weak solids viewed from the jammed phase?

In this letter, we show that acoustic surface waves sounding provides new insight into the structure of the elastic networks in the vicinity of jamming. This experimental technique allows to measure accurately the elastic properties of a granular packing under vanishing pressure --~namely, close to a free surface, under gravity loading. In ordinary elastic solids, surface waves (called Rayleigh waves) are a combination of compression and shear waves and travel at a speed slightly smaller than the bulk shear wave~\cite{LL86} ($\sim 5000~$m/s for glass). In the context of granular matter, ethologists have reported that numerous species living at the desert surface use sound waves to probe their environement. Indeed, the first experiments of surface mode propagation were conducted by biologists who interpreted their signal as a collection of fast (bulk) waves and slow (shear) waves although the measured speeds ($\sim 50~$m/s) were incredibly small~\cite{B77}. This was confirmed by field measurements performed at the surface of sand dunes in the context of booming avalanches studies~\cite{A04,ABC08}. Recently, these surface waves were ascribed to  the gravity induced index gradient, which plays the role of a wave-guide,  and analysed in two slightly different models of non-linear elasticity~\cite{G06a,BAC07}. 
\begin{figure}[t!]
\includegraphics{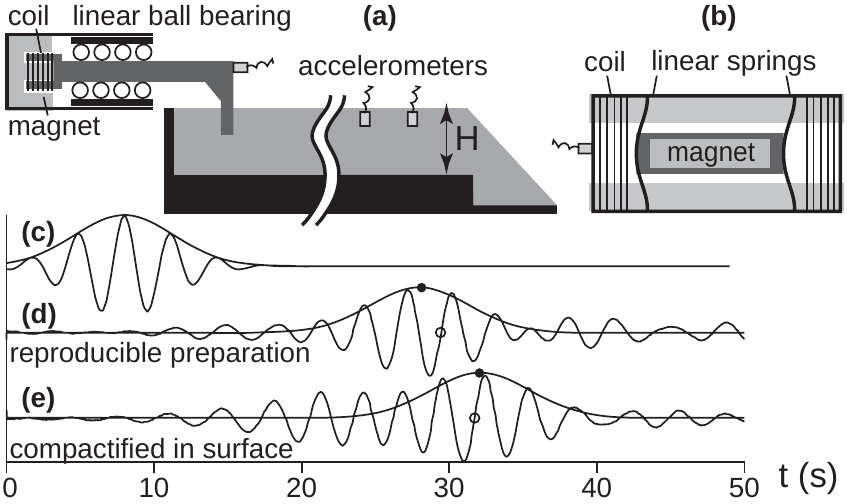}
\vspace{-0mm}
\caption{{\bf (a)} Experimental set-up. Acoustic source of sagital waves. {\bf (b)} Acoustic source of transverse waves. {\bf (c)} Signal of the accelerometer mounted on the source.   {\bf (d-e)}  Transverse wave-packets received at $x=60~cm$ from the source, for two preparations of the granular packing (see text). Note the shift of the wave-packet centre ($\bullet$) and of the phase $\circ$.} 
\vspace{-0mm}
\label{setup} 
\end{figure}

To clarify the theoretical issues addressed here, let us rephrase in the framework of non-linear elasticity the predictions of Wyart et al.~\cite{WSNW05} on the elastic anomaly induced by soft-modes. For geometrical reasons, the Hertz contact force between two grains depends non-linearly on their relative inter-penetration $\delta$. On this basis, the macroscopic elastic free energy~\cite{LL86} of an isotropic granular packing can be written in the very general form~\cite{JL03}: 
\begin{equation}
\mathcal{F}=E  \left(\frac{2}{5} \mathcal{B} \delta ^{5/2}+ \mathcal{A} \delta ^{1/2} u_{ij}^{0} u_{ij}^{0}\right)
\label{free_energy}
\end{equation}
where $u_{ij}$ is the coarse-grained strain tensor, $\delta=-{\rm Tr}(u_{ij})$ is the volumic compression and $u_{ij}^{0}=u_{ij}+\frac{\delta}{3} \delta_{ij}$ the traceless strain tensor. $\mathcal{A}$ and $\mathcal{B}$ are two dimensionless elastic coefficients that characterise the material stiffness under shear and compression, respectively. We assume that the average number of contacts per grain $\mathcal{Z}$ is sufficient to characterise the microscopic packing geometry and thus, that $\mathcal{A}$ and $\mathcal{B}$ are function of $\mathcal{Z}$. Indeed, in a frictional packing, different values of   $\mathcal{Z}$ can be obtained under the same pressure $p$; $\mathcal{Z}$  and $p$ are thus independent state variables \cite{MRJWM08}. It should be emphasised that $\mathcal{F}$ is not supposed to describe the stress-strain curve obtained from a loading test, which is composed by series of elastic loadings at fixed $\mathcal{Z}$ and of plastic events during which $\mathcal{Z}$ changes.

The mean field calculation~\cite{BAC07} predicts that $\mathcal{A}$ and $\mathcal{B}$ vary linearly in $\mathcal{Z}$ and thus remain finite at $p=0$. On the other hand, the jamming theory predicts that upon approaching isostaticity, the stiffness ratio $\mathcal{A}/\mathcal{B}$ exhibits an anomalous behaviour --~as well as many other quantities~-- and scales on the excess of contacts above the isostatic value, $\mathcal{Z}-\mathcal{Z}_{\rm iso} \propto \delta^{1/2}$. So, this approach predicts that $\mathcal{A}/\mathcal{B}$ vanishes with pressure. Let us emphasise again that the reality of the jamming point as a critical point was assessed for frictionless soft spheres but there are several indications that this line of ideas could be generalised to frictional packings~\cite{SHESS07,SHS07} and even to glasses~\cite{XWLN07}.

Starting from expression (\ref{free_energy}), the surface elastic modes can be derived as shown in~\cite{BAC07}. They propagate through an infinite though discrete collection of modes polarized sagitally or transversally. The mode labelled $n$ of wavelength $\lambda$ penetrates the sample over a typical depth $n\lambda$ and thus feels a typical pressure $p=\rho g n \lambda$. Then, the mean field theory predicts a velocity dependence of the form: $v\sim (n\,\lambda)^{1/6}$ while the "soft modes" theory predicts, instead, $v\sim (n\,\lambda)^{1/3}$. Thus, the dispersion relation of surface waves constitutes a direct experimental test of the existence of a shear stiffness anomalous scaling.
\begin{figure}[t!]
\includegraphics{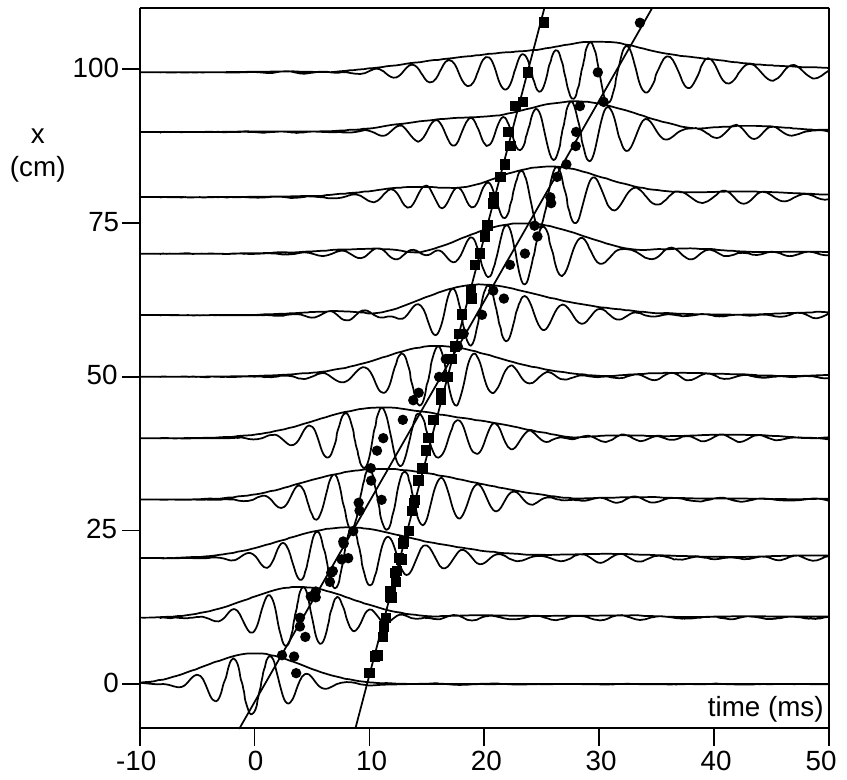}
\vspace{-0mm}
\caption{Space time diagram showing the wave-packet propagation. Starting from the raw signals received at different positions, the wave-packet is roughly localized by computing the signal envelope. Then, a local fit by a gaussian wave-packet allows to determine the centre of the wave-packet ($\bullet$) and its phase with respect to the source (symbols $\blacksquare$ show the space-time coordinates of an iso-phase event). The best fit (thin lines) allows to extract the group and phase velocities.} 
\vspace{-0mm}
\label{xoft} 
\end{figure}
 
As confirmed by preliminary experiments~\cite{ABC08}, the weak dependence of the speed of sound on $n$ makes the experiment very difficult to control and almost impossible to analyse. Indeed, using any standard source, a huge number of propagating modes are excited that remain superimposed over large distances. To bypass this problem, we have successfully designed an experiment that isolates the first sagittal and transverse modes. For this, measurements are performed  in a rectangular channel of width $W=20~$cm and length $180$~cm, which serves as a second wave-guide (Fig.~\ref{setup}a): the lateral boundary conditions impose a relation between the measured wavelength $\lambda$ and the wavelength $\lambda_\infty$ that would be selected in an infinitely large channel, at the same frequency:
\begin{equation}
\lambda_\infty=\left[\frac{1}{\lambda^{2}}+\left(\frac{m}{2W}\right)^2\right]^{-1/2}
\end{equation}
where $m$ is the transverse mode number ($m=1$ here). Moreover, the sources are conceived and tuned to excite essentially the modes $n=1$: sagittal waves are produced by an  electromagnetic shaker (without any spring) whose axis is finely guided by a ball bearing slider coupled to a very rigid transverse metallic blade;  transverse waves are produced by a  rough cylinder inside which a permanent magnet  vibrates under the action of a magnetic field (fig.~\ref{setup}b). The channel is filled of glass beads ($E=70$~GPa, $d=150~\mu$m) over the  height $H=20~$cm. For such an aspect ratio, with smooth boundaries, the Janssen effect is negligible so that the pressure is expected to vary linearly in depth. The acoustic isolation is insured by $20$~cm thick boundaries.
 
The experiment is also designed to prevent another problem. As the sample presents random heterogeneities, the acoustic signal is composed by an effective medium response and a coda related to speckle effect~\cite{JCV99}. Their relative amplitude is controlled by the number of grains in contact with the transducer. We have chosen to work with accelerometers of diameter $D=13$~mm which allow to measure the three components of the acceleration in the bulk of the sample. Around $3~10^4$ grains are in contact with the transducer and the measured amplitude of the coda-tail is around $5\%$ of the coherent signal. By comparison, other techniques like a laser vibrometer would only probe the rough surface of the packing and due to the small size of the spot ($5~$mm) would only average over $\sim 400$ grains, yielding a coda and a coherent signal of the same order of magnitude. We have also checked that the propagation is not affected by the presence of other accelerometers between the source and the receiver. Besides, the transducer size $D$ should be at least a fraction of the wavelength $\lambda$, which imposes to work at rather low frequencies $f<1~$kHz. In summary, the experiment has to be analysed having in mind the hierarchy of length scales: $d \ll D<\lambda<H$.  
\begin{figure}[t!]
\includegraphics{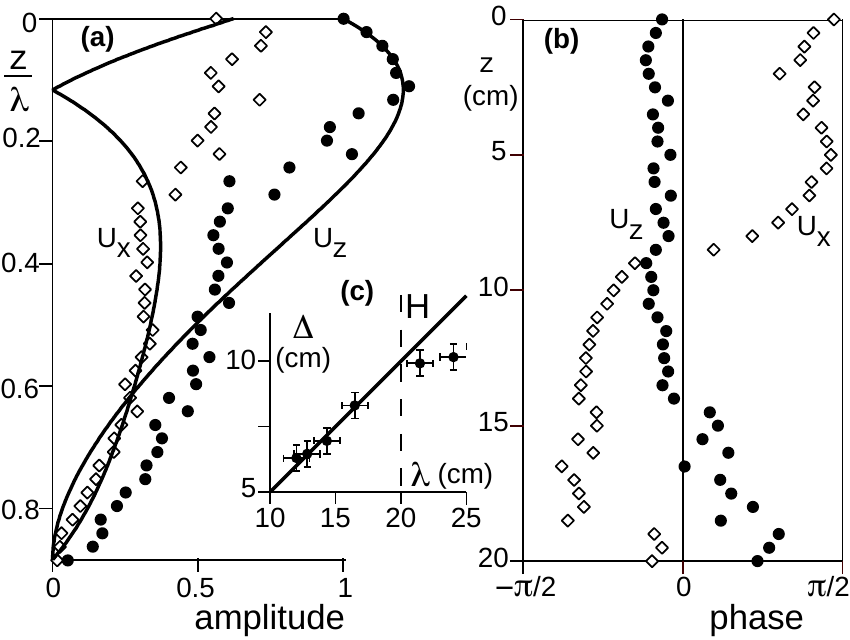}
\vspace{-0mm}
\caption{First sagittal mode. {\bf (a)} Amplitude and {\bf (b)} phase of the vertical ($\bullet$) and horizontal ($\diamond$) displacement as a function of depth, for $f=315$~Hz ($\lambda \simeq 21.5$~cm). {\bf (c)} Length $\Delta$ over which the vibration decays as a function of the wavelength $\lambda$.} 
\vspace{-0mm}
\label{profile} 
\end{figure}

The typical vibration amplitude we use is $\sim 10~$nm, although the propagation properties remain the same up to $\sim 100~$nm i.e. a strain of $10^{-6}$. Above, new peaks appear in the signal and period doubling is observed in the coda. The preparation of the sample is amongst the most difficult part of the experiment. Prior to each measurement, we systematically sweep a thin blade longitudinally and transversally through the packing in order to remove any memory effect due to the granular initial filling or the subsequent accelerometers manipulation. The extra-sand above the level of the channel is then gently removed, letting a flat surface. The figure~\ref{setup} compares a typical signal obtained with this procedure (d) to that obtained when pouring the grains and flattening the surface by tapping with a hammer on a plaster float (e). In the first case, the phase and the travel-time of the wave-packet is reproducible. By contrast, the apparent phase velocity and group velocity can vary up to $25\%$ from one sample to the other, although they are macroscopically identical. This non-universality is the first important conclusion of this letter:  the packing elastic properties depend on the the preparation protocol.

Fig.~\ref{xoft} shows the positions of the centre of the wave-packet and of an iso-phase event as a function of time. For this, we prepared $\sim 20$ realisations of the packing. For each, the signals of four accelerometers placed at different positions are acquired and fitted by a gaussian wave-packet. The propagation of a coherent mode --~and not speckle~-- is clearly evidenced by the linear relationship between space and time and by the reproducibility over independent microscopic realisations. The slopes of the relations respectively give the group and the phase velocities, $v_g$ and $v_\phi$. Looking now in depth, the vibration amplitude is observed to decrease over a distance $\Delta$, of the order of a half wavelength (fig.~\ref{profile}c). The sagittal waves are elliptically polarized, with their principal axis in the vertical direction and along the direction of propagation. The figure~\ref{profile} compares the shape of the first mode to the prediction of our model~\cite{BAC07}. We obtain a fair agreement with the theory which confirms that the first mode has been successfully isolated, as desired. 
\begin{figure*}[t!]
\includegraphics{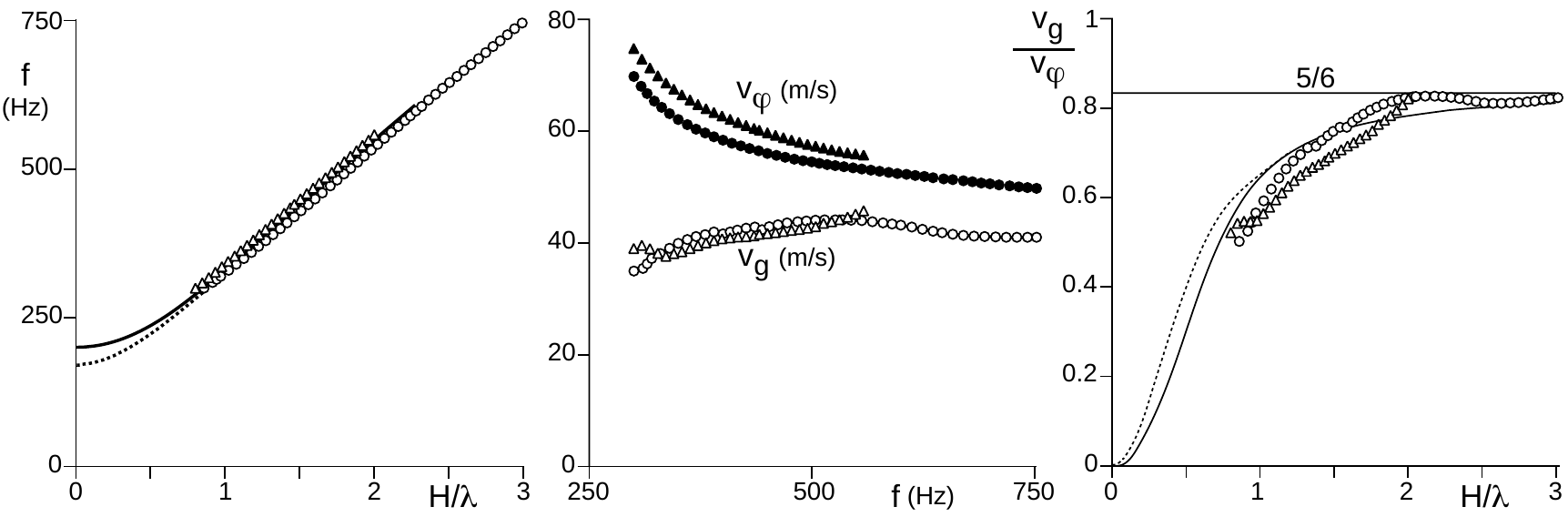}
\vspace{-0mm}
\caption{{\bf (a)} Dispersion relation of sagittal ($\circ$) and longitudinal waves ($\triangle$):  frequency $f$ as a function of the rescaled wavenumber $H/\lambda$.  {\bf (b)}  Corresponding group $v_g$ (white symbols) and phase $v_\phi$ (black symbols) velocities. {\bf (c)} Ratio of the group and phase velocities $v_g/v_\phi$ as a function of $H/\lambda$. In the three graphs, only one fifth of the measured points are shown. The predictions of the model, including the finite width and depth, are shown in solid (sagittal) and dotted (transverse) lines.} 
\vspace{-0mm}
\label{dispersion} 
\end{figure*}

In our set-up, the working frequency $f$ can be varied in the range $300-750$~Hz  for sagittal waves, and $300-550$~Hz for transverse waves.  Outside these frequency ranges, the signals received are either distorted due to the source, or multi-modal. No propagation at all is observed below $200$~Hz. We have measured $v_g$ and $v_\phi$, as well as their statistical uncertainty, every $3$~Hz. These values are used to reconstruct accurately a single dispersion relation $f(\lambda^{-1})$ (fig.~\ref{dispersion}) that simultaneously fits  the group and phase velocities in the least square sense. As expected, the propagation is dispersive since  $v_g$ and $v_\phi$ are different. This can be related to the two wave-guide effects previously mentioned. Due to the finite depth and width, a cut-off frequency appears in the dispersion relation, which corresponds to the first resonance of the system. The group velocity is expected to vanish at this frequency and the phase velocity to diverge. This explains the behaviour observed at low frequency, and in particular the increase with $f$ of the ratio $v_g/v_\phi$. At high frequency, an asymptotic behaviour controlled by the pressure induced wave-guide is reached in the limit of wavelengths $\lambda$ small in front of the channel transverse dimensions ($H$ and $W$). The ratio $v_g/v_\phi$ tends to a constant equal to $0.82\pm0.04$ for the preparation described above.  As $v_g$ is the slope of the dispersion relation, $v_g/v_\phi$ is the scaling exponent between $f$ and $\lambda^{-1}$. The measured value is very close to that expected if $\mathcal{A}$ does not vanish at the surface ($5/6$). Thus, our experimental results fully confirm the Hertzian picture down to $\lambda \sim 250\,d$ and does not show any evidence of anomalous exponent ($2/3$) when $p \rightarrow 0$, that would be associated to the jamming point~\cite{WSNW05,SHS07}. It would be interesting to pursue this technique to lower the probing wavelengths in order to push the limits to weaker confining pressures. But to this purpose, the limitations due to finite probe size, preparation sensitivity and speckle noise have to be overcome. 

Globally, one can compare the experimental data to the prediction of the model, assuming that $\mathcal{A}$ does not vanish at the surface and remains nearly constant at the scale $\lambda$ and taking into account the finite width and depth (lines on fig.~\ref{dispersion}). The agreement  is excellent (within $5\%$). The dispersion relations of transverse and sagittal modes turn out to be nearly equal (fig.~\ref{dispersion}). This striking behaviour is one of the robust output of our model~\cite{BAC07}. For ratios $\mathcal{B}/\mathcal{A}=O(1)$, the difference would only be of  $10\%$ and would be almost indistinguishable when, say, $\mathcal{A}<0.2~\mathcal{B}$. The physical reason is that the restoring force for both modes is the shear elasticity (parameter $\mathcal{A}$). In the limit $\lambda \ll H$ the dispersion relation of the first modes  takes the form:
\begin{equation}
f = \alpha (E/\rho)^{1/3} g^{1/6}  \lambda_\infty^{-5/6}\,\,{\rm with}\,\,\alpha \simeq 0.77 \, \mathcal{A}^{1/2} \mathcal{B}^{-1/6} 
\end{equation}
The best fit gives: $f \lambda_\infty^{5/6}\simeq 77\pm1~{\rm m}^{5/6} {\rm s}^{-1}$, which corresponds to a value of $\mathcal{A}^{1/2} \mathcal{B}^{-1/6} \sim 0.23$. By contrast, the mean-field expectation is $0.40$ for frictionless grains and $0.61$ for infinite friction.  Thus, the measured shear stiffness is $3$ to $5$ times smaller than expected from the mean field theory, as observed in numerical simulations \cite{MGJS99}. The "soft-modes" theory is the only one explaining this mean-field failure. Still, to be consistent with our results, one has to conclude that the packing does not tend to isostaticity in surface: for frictional packings prepared in a simple way, $\mathcal{Z}$ remains significantly larger than $\mathcal{Z}_{\rm iso}$ under vanishing pressure. 

We thank D. Pradal, D. Renard, T. Darnige and J. Lanuza for their technical assistance and P. Claudin for his critical reading of the manuscript.


\begin{thebibliography}{}
\bibitem{TWLB02} A.~Tanguy, J.P. Wittmer, F. Leonforte and J.-L. Barrat, Phys.Rev. B \textbf{66}, 174205 (2002).
\bibitem{OHSLN03} C. S. O'Hern, L. E. Silbert, A. J. Liu and S. R. Nagel, Phys. Rev. E \textbf{68}, 011306 (2003).
\bibitem{WSNW05} M.Wyart, L. E. Silbert, S. R. Nagel, and T. A. Witten, Phys. Rev. E \textbf{72}, 051306 (2005).
\bibitem{SHESS07} E. Somfai, M. van Hecke, W.G. Ellenbroek, K. Shundyak, and W. van Saarloos Phys. Rev. E \textbf{75}, 020301 (R) (2007)
\bibitem{MRJWM08} V. Magnanimo, L. La Ragione, J. T. Jenkins, P. Wang and H. A. Makse, EPL \textbf{81}, 34006 (2008). 
\bibitem{LL86} L.D.~Landau and E.M.~Lifshitz, \textit{Theory of elasticity} New York Pergamon Press 3rd edition (1986).
\bibitem{B77} P.H.~Brownell, Science \textbf{197}, 479-482 (1977); Sci. Am. \textbf{251}(6) 94-105 (1984); P.H.~Brownell and R.D.~Farley, J. Comp. Physiol. \textbf{131}, 23-30, 31-38 (1979).
\bibitem{A04} B.~Andreotti, Phys. Rev. Lett. \textbf{93}, 238001 (2004).
\bibitem{ABC08} B. Andreotti, L. Bonneau, and E. Cl\'ement, accepted to J. Geo. Res., arXiv:0706.0263 (2008).
\bibitem{G06a} V.~Gusev, V.~Aleshin, V.~Tournat., Phys. Rev. Lett. \textbf{96}, 214301 (2006).
\bibitem{BAC07} L. Bonneau, B. Andreotti, and E. Cl\'ement Phys. Rev. E \textbf{75}, 016602 (2007).
\bibitem{JCV99} X. Jia, C. Caroli and B. Velicky, Phys. Rev. Lett. \textbf{82}, 1863 (1999).
\bibitem{MGJS99} H.A.~Makse, N.~Gland, D.L.~Johnson and L.M.~Schwartz, Phys. Rev. Lett. \textbf{83}, 5070 (1999) and Phys. Rev. E \textbf{70}, 061302 (2004).
\bibitem{JL03} Y.~Jiang and M.~Liu, Phys. Rev. Lett. \textbf{91}, 144301 (2003); \textbf{93}, 148001 (2004).
\bibitem{SHS07} K. Shundyak , M. van Hecke  and W. van Saarloos , Phys. Rev. E \textbf{75},  010301 (2007).
\bibitem{XWLN07}  N. Xu , M. Wyart , A.J. Liu and S.R. Nagel , Phys. Rev. Lett. \textbf{98}, 175502 (2007).
\end{thebibliography}
\end{document}